\documentclass[%
aip,
jcp,%
amsmath,amssymb,
reprint,%
]{revtex4-1}

\usepackage{color}
\usepackage{yfonts}
\usepackage[pdftex]{graphicx}
\setkeys{Gin}{clip=true,draft=false}
\DeclareGraphicsExtensions{.pdf}
\usepackage{amsthm}
\usepackage{amsfonts}
\usepackage{amssymb}
\usepackage[cspex,bbgreekl]{mathbbol}
\usepackage[hang]{subfigure}
\usepackage{float}
\usepackage[super]{nth}
\usepackage{multirow} 
\usepackage{tabulary}
\newcolumntype{K}[1]{>{\centering\arraybackslash}p{#1}}
\newtheoremstyle{definition_new}
{0.7\topsep}
{0.7\topsep}
{\normalfont}
{0pt}
{\itshape}
{}
{7pt}
{}
\theoremstyle{definition_new}

\theoremstyle{remark}

\usepackage{graphicx}
\usepackage{dcolumn}
\usepackage{bm}
\usepackage[english]{babel}

\begin{document}
	
	
\title{Chaos or Order?}

\author{Igor V. Ovchinnikov} 
\email{email: iovchinnikov@ucla.edu}
\affiliation{Electrical Engineering Department, University of California at Los Angeles, Los Angeles, 90095 CA}
	
\author{Massimiliano Di Ventra}
\email{email: diventra@physics.ucsd.edu}
\affiliation{Department of Physics, University of California, San Diego, La Jolla, 92093 CA}

\begin{abstract}
What is chaos? Despite several decades of research on this ubiquitous and fundamental phenomenon there is yet no agreed-upon answer to this question. Recently, it was realized that all stochastic and deterministic differential equations, describing all natural and engineered dynamical systems, possess a topological supersymmetry. It was then suggested that its spontaneous breakdown could be interpreted as the stochastic generalization of deterministic chaos. This 
conclusion stems from the fact that such phenomenon encompasses features that are traditionally associated with chaotic dynamics such as non-integrability, positive topological entropy, sensitivity to initial conditions, and the Poincar\`e-Bendixson theorem. Here, we strengthen and complete this picture by showing that the hallmarks of set-theoretic chaos -- topological transitivity/mixing and dense periodic orbits -- can also be attributed to the spontaneous breakdown of topological supersymmetry. We also demonstrate that these features, which highlight the ``noisy'' character of chaotic dynamics, do not actually admit a stochastic generalization. We therefore conclude that spontaneous topological symmetry breaking can be considered as the most general definition of continuous-time dynamical chaos. Contrary to the common perception and semantics of the word “chaos”, this phenomenon should then be truly interpreted as the low-symmetry, or \emph{ordered phase} of the dynamical systems that manifest it. 
Since the long-range order in this case is temporal, we then suggest the word ``chronotaxis'' as a better representation of this phenomenon. 
\end{abstract}

\maketitle

The word ``chaos'' was first introduced by Li and Yorke~\cite{LiYorke} to describe certain dynamical systems and to highlight that even deterministic dynamics may be, at sufficiently long times, highly irregular and unpredictable, a fact that was already known from the early computational work of Lorenz on modeling atmospheric convection~\cite{Lorenz}, and can be traced back to pioneering studies in the late 19th century by Hadamard~\cite{Hadamard} and Poincar\'e~\cite{Poincare} on Hamiltonian dynamics. 

Although ``irregularity'' and ``unpredictability'' may seem difficult concepts to quantify, mathematicians have provided some rigorous definitions of chaotic dynamics that make precise some of its characteristics. For instance, the word ``chaos'' is sometimes considered synonym of positive ``topological entropy''~\cite{Adler}, where the latter is a non-negative real number (possibly infinite) that quantifies the ``complexity'' of the flow~\cite{Manning}. In fact, a conjecture was advanced by Shub~\cite{Shub}, later proved by Yomdin~\cite{Yomdin} for $C^{\infty}$-smooth maps on compact spaces, that the spectral radius of the pushforward map of the associated homology groups provides a lower bound for the topological entropy. 

In more modern approaches, a deterministic dynamical system defined by a map $M: X \to X$ of a compact phase space $X$ onto itself, is typically identified as chaotic if it exists an invariant compact subset, $Y\subset X, M(Y)=Y$, such that the following three properties are satisfied~\cite{Devaney,newpaper}: 
\emph{(i)} The periodic orbits of $M$ are dense in $Y$,
\emph{(ii)} $M$ is topologically transitive on $Y$, and \emph{(iii)} $M$ depends sensitively on initial conditions on $Y$. Topological transitivity means that, given any two points in $Y$, we can find an orbit that comes arbitrarily close to both.

The last property \emph{(iii)} can be quantified by calculating the Lyapunov spectrum and showing that at least one Lyapunov exponent is positive. Such positive Lyapunov exponent provides a measure of the degree of instability of the system~\cite{Vulpiani}. 

Property \emph{(iii)} is indeed redundant -- and alone is typically not sufficient to characterize a system as chaotic~\cite{Katok} -- since one can prove~\cite{Banks} that if the dynamical system satisfies \emph{(i)} and \emph{(ii)}, then \emph{(iii)} follows~\cite{Banks}~\footnote{In this work we do not restrict ourselves to continuous maps on intervals of $\mathbb{R}$ for which transitivity implies also a dense set of periodic orbits~\cite{Vellekoop}.}. 

Finally, transitivity can be replaced by the much stronger condition of {\it topological mixing}~\cite{book-top-dyn}, which means that any open set of $Y$ overlaps with any other open set if the former is propagated long enough. Topological mixing implies transitivity, but the opposite is not generally true.  

The ensemble of the above features --topological transitivity/mixing, dense orbits-- characterize what can be called {\it set-theoretic chaos}. These features are universal in the sense that they hold for both step-like (discrete-time) and continuous-time dynamics. 

For continuous-time dynamical systems, there is yet another widely accepted definition of chaos~\cite{Gilmore}. This definition is the {\it non-integrability} of the flow vector field that was first noted by Poincar\'e~\cite{Poincare}. Non-integrable flows are such that their {\it global} submanifolds of the phase space are not well defined topological manifolds. Rather, they can be qualitatively represented as ``branched manifolds'', formalizing the concept of ``folding'' and ``stretching'' in non-integrable continuous-time dynamics~\cite{Gilmore}. In fact, for continuous-time dynamics, non-integrability alone is sufficient to characterize a system as chaotic. Finally, phase-space constraints lead to the Poincar\'e-Bendixon theorem, which states that continuous-time, deterministic dynamical systems can only be chaotic if the dimensionality of the phase space is higher than two (see, e.g.,~\cite{Vulpiani}).

There are quite a few relations between all these conditions. Some topological and metric features can be used together to provide strong evidence of chaotic behavior. For instance, if the system is both topologically transitive (global property) and has positive topological entropy (metric property) then (\emph{iii}) follows~\cite{Katok}. It is not our intention to review all of these relations. Rather, we seek a {\it unifying principle} that encompasses all of them. 

All these characteristics, although apparently distinct, tend to highlight one aspect that is now well accepted and unquestioned by the scientific community: chaos is interpreted as a spontaneously emerging ``disorder'' or ``noise'' of the associated dynamical system. This interpretation is somewhat suspicious in light of property \emph{(iii)}, known also as the ''butterfly effect''~\cite{Lorenz}. This can be understood as a spontaneous, infinitely-long memory of perturbations. The concept of ``disorder'' is clearly in contradiction with the concept of long-term memory. 

This seeming contradiction may be resolved as follows. In Physics, a ``disordered phase'' is synonymous to ``symmetric phase'', namely it has more symmetries than its ordered counterpart, which is instead the {\it low-symmetry} phase of the system. Consider, for instance, a liquid vs. a crystal: the disordered or symmetric liquid phase has both translational and rotational symmetries, while the ordered or lower-symmetry crystal does not. Similarly, the ordered ferromagnetic phase of a solid does not possess the $SU(2)$ spin rotation symmetry of the paramagnetic phase, where the spins are randomly oriented, and so on.

However, it does not seem straightforward to interpret chaos as a phenomenon emerging from symmetry breaking. Chaos occurs in such a diverse class of dynamical systems~\cite{Vulpiani}, that it seems difficult to identify a set of symmetries, or even one symmetry, common to all of these systems and encompassing all the properties we have discussed above. 

Continuous-time dynamical systems offer such an interpretation. Recently, it was shown that {\it any} continuous-time dynamical system described by stochastic or  
deterministic (partial) differential equations possesses topological (de Rahm) supersymmetry and that the stochastic generalization of chaos can be identified with the spontaneous breakdown of such symmetry~\cite{Entropy,Chaos}. This idea has clarified the physical origin of self-organized criticality \cite{STS_chapter}, and the presence of topological supersymmetry in all dynamical systems has found an important application in understanding how certain engineered dynamical systems, known as memcomputing machines~\cite{UMM}, can solve complex problems efficiently~\cite{Di_Ventra}.

The goal of this paper is to strengthen and complete the topological supersymmetry breaking (TSB) picture of dynamical chaos by focusing on set-theoretic chaos (STC). We will first restate in Section~\ref{SecI} the results on the stochastic case~\cite{Entropy,Chaos}. This will allow us to introduce the key ingredients of the TSB picture and discuss how it provides stochastic generalizations to non-integrability in the sense of dynamical systems, positive topological entropy, sensitivity to initial conditions, and Poincar\'e-Bendixson theorem. We will then move on to the deterministic limit in Section ~\ref{Sec:II}, where we discuss the relation between  
TSB and STC. We will show that STC, together with its trajectory-based properties, does not admit a stochastic generalization. Therefore, both from the 
perspective of universality and from a physical point of view, TSB offers the most general definition of chaos and the most physical one because there is no system in Nature that is truly noiseless. We then conclude that if TSB is accepted as a universal definition of continuous-time chaotic dynamics, both deterministic and stochastic, then this phenomenon should be truly interpreted as the low-symmetry, {\it ordered phase} of the dynamical systems that manifest it. Since the ensuing long-range order is 
temporal, we think the word {\it chronotaxis} better represents this phenomenon. 


\section{Topological supersymmetry breaking}\label{SecI}


We summarize here the main ingredients of the supersymmetric theory of stochastic dynamics that will be needed later. A more detailed review of the subject can be found in Ref.~\cite{Entropy}. 
This theory employs concepts that have counterparts in Quantum Mechanics (QM), but are specific to the theory of classical dynamical systems. The advantage of this is precisely that many results can be extracted by working, as in QM, in a Hilbert space of states, and 
using an evolution operator. However, the major difference with QM is that the evolution operator in the present theory is pseudo-Hermitian~\cite{Mos023}.

Let us start by considering a generic continuous-time dynamical system with noise. It can be defined via a stochastic differential equation (SDE) of the type
\begin{equation}
	\dot x(t) = F(x(t))+(2\Theta)^{1/2}e_a(x(t))\xi^a(t), \label{SDE}
\end{equation}
with $x\in X$ being the position in the D-dimensional topological manifold called phase space~\footnote{In the spirit of dynamical systems theory, one can think of $X$ as an infinite metric compact space with Hausdorff topology.}, $F\in TX$ being the flow vector field (on the tangent space $TX$) representing the deterministic laws of temporal evolution, 
$\Theta$ is the intensity of the noise, $e_a$ are vector fields that represent the coupling of the noise to the system, and $\xi\in \mathbb{R}^D$ is a Gaussian white noise~\footnote{This choice of noise is by no means necessary: any other physical noise would be equally valid.}, defined via the probability of configurations,
\begin{eqnarray}
	P(\xi) \propto e^{-\int dt \xi^2(t)/2}.\label{Noise_Probability}
\end{eqnarray}

The mathematical benefit of adding noise is to render the stochastic evolution operator (SEO) -- we write explicitly in Eq.~(\ref{H_SEO}) below-- elliptic, with a spectrum bounded from below. The physical reason we have added noise is that in the real world, no dynamics are exempt from it. In the deterministic limit, of zero noise, $\Theta \rightarrow 0$, the 
stochastic evolution operator, Eq.~(\ref{H_SEO}), may lose its ellipticity, and possibly even give an unbounded (from below) spectrum. However, as we will explain later, this does not invalidate the picture of the topological supersymmetry breaking in the deterministic case. 

Eq.~(\ref{SDE}) defines the family of noise-configuration-dependent maps, $M_{tt'}:X\to X$, defining the trajectories as $x(t) = M_{tt'}(x(t'))$. For any configuration of noise, even for non-continuous ones, $M_{tt'}$ is a diffeomorphism as established~\cite{Slavik2013261} using Kurzweil's theory~\cite{Kurzweil}.

Now, we define the finite-time stochastic evolution operator (SEO), the adaptation of the generalized transfer operator of dynamical systems theory~\cite{Ruelle} to SDEs,
\begin{eqnarray}
	\hat {\mathcal M}_{tt'}=\langle M^*_{t't}\rangle,\label{SEO}
\end{eqnarray}
where angled brackets denote stochastic averaging over all configurations of noise, and  $M^*_{t't}:\Omega(X)\to\Omega(X)$ is the action or pullback induced by $M_{t't}$ on the exterior algebra of $X$, $\Omega(X)$, considered as the Hilbert space of the system and its elements --the differential forms of various degrees-- viewed as wavefunctions of the model. Note that even though we use the term wavefunctions for the elements of $\Omega$, they are not wavefunctions in the sense of QM: they are the objects that represent possible states of the model. For example, top differential forms represent the total probability distributions over $X$ in a coordinate-free setting.

With a little math, the finite-time SEO in Eq.~(\ref{SEO}) is found to be~\cite{Entropy},
\begin{equation}
	\hat{\mathcal{M}}_{tt'} = e^{- \hat H (t-t')},\;\;\;\;\;\; \hat H = \hat L_F - \Theta \hat L_{e_a}\hat L_{e_a},\label{H_SEO}
\end{equation}
where $\hat L_F = [\hat d, \hat \imath_F]$ is the Lie derivative with $\hat \imath_F$ being the so-called interior multiplication along $F$, and $\hat d$ is the nilpotent ($\hat d^2=0$) exterior derivative or de Rahm operator~\cite{Topo-book}, and square brackets denote the bi-graded commutation, i.e., the anti-commutation if both operators are ``fermionic'' (have odd degrees in $\Omega$), and the commutation otherwise. The meaning of Lie derivative is the infinitesimal pullback which explains why it enters into the expression for the infinitesimal SEO, $\hat H$ (not to be confused with the Hamiltonian in QM), in Eq.(\ref{H_SEO}). 

The SEO $\hat H$ is a real operator and thus it is pseudo-Hermitian~\cite{Mos023}. This means that its eigenvalues are either real or come in complex conjugate pairs\footnote{These pairs can be thought of as stochastic generalizations of Reulle-Pollicott resonances of dynamical systems theory.} and that its (bi-orthogonal) eigensystem is complete so that any differential form (wavefunction), $|i\rangle\in\Omega$, can be resolved as,  
\begin{eqnarray}
	|i\rangle=  \sum\nolimits_{n} C_{ni}|n\rangle,
\end{eqnarray} 
where $C_{ni} = \langle n | i \rangle $ are the overlapping coefficients. The left and right eigenstates of $\hat H$ (analogous to the bras and kets in QM) are defined by,
\begin{equation}
	\langle n | \hat H = \langle n | \mathcal{E}_n, \,\,\,
	\hat H |n\rangle  = \mathcal{E}_n |n \rangle,\label{eigensystem}
\end{equation}
with $\mathcal{E}_n$ being the corresponding eigenvalue and $\langle n|k\rangle = \delta_{nk}$. The relation between left and right eigenstates is more complicated than that in QM and $\langle n| \ne (|n\rangle)^*$ in the general case. 
\begin{figure}
	\begin{centering}
		\includegraphics[width=8cm]{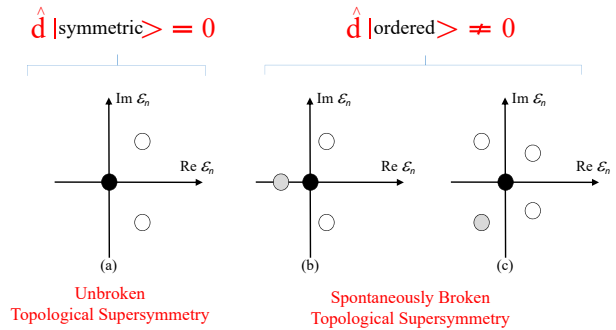}
		\caption{\label{Fig_Spectrum} Qualitative picture of the three possible different types of spectra of the infinitesimal SEO~(\ref{H_SEO}). The eigenstates with the lowest real part of their eigenvalues contribute the most to the partition function and are thus the ground states. (a) The supersymmetry is unbroken because the ground states are zero-eigenvalue supersymmetric states (the $\hat d$ operator annihilates all of them). One of them (black filled dot) is the steady-state total probability distribution known as ergodic zero or thermal equilibrium. (b)-(c) Supersymmetry is spontaneously broken because the ground states (gray filled dots) have non-zero eigenvalues and are thus non-supersymmetric (the $\hat d$ operator does not annihilate all of them). In (c), there are two equally good candidates for the status of ground state. The one on the bottom (gray dot) can be chosen by convention, similarly to how the ground state is chosen in QM.}
	\end{centering}
\end{figure}

With the help of the SEO~(\ref{SEO}), the temporal evolution of any differential form is 
\begin{eqnarray}
	|i,t\rangle = \hat{\mathcal{M}}_{tt'} |i\rangle=  \sum\nolimits_{n} e^{-(t-t')\mathcal{E}_n}C_{ni}|n\rangle.
\end{eqnarray}
This shows that the original nonlinear trajectories in $X$ are now described by a linear evolution on $\Omega(X)$. This ``linearization'' comes at a price: the dimensionality of $\Omega(X)$ is infinitely larger than $\textrm{dim} X = D$.

So far, the theory can be thought of as a generalization of the Fokker-Planck evolution of the total probability distributions (top differential forms) to differential forms of all possible degrees. This generalization in known in dynamical systems theory as the generalized transfer operator formalism ~\cite{Ruelle}. The qualitative step forward happens when one notices that the exterior derivative commutes with the SEO,
\begin{equation}
	[\hat d, \hat H] = 0,\label{commute} 
\end{equation}
hence, it is a \emph{supersymmetry} of the model. This means that if $|n\rangle$ is an eigenstate with eigenvalue, $\mathcal{E}_n$, so is $\hat d|n\rangle$,
\begin{equation}
	\hat d \hat H |n\rangle  = \mathcal{E}_n \hat d |n\rangle \to \hat H \hat d |n \rangle = \mathcal{E}_n \hat d |n\rangle.\label{eigensystem}
\end{equation}
In this manner, almost all the eigenstates are paired up into non-supersymmetric doublets, $|n\rangle$ and $|n'\rangle = \hat d | n\rangle$. Only some eigenstates are supersymmetric singlets, $|\alpha \rangle$, that are non-trivial in de Rahm cohomology, i.e., they are such that $\hat d | \alpha \rangle=0$ but no state $|\tilde\alpha\rangle$ exists such that $|\alpha\rangle = \hat d |\tilde \alpha\rangle$. An equivalent definition of supersymmetric states is the requirement that the expectation value of any $\hat d$-exact operator, i.e., of the form $[\hat d, \hat X]$ (with $\hat X$ an arbitrary operator), vanishes,
\begin{eqnarray}
	\langle \alpha | [\hat d, \hat X] | \alpha \rangle = 0, \;\;\;\forall \hat X.
\end{eqnarray}
Note that $\hat H$ is a $\hat d$-exact operator,
\begin{equation}
	\hat H = [\hat d, \hat{\bar d}],\label{DExactFPOperator}
\end{equation}
where $\hat{\bar d} = F^i\hat{\imath}_i - \Theta e^i_a\hat{\imath}_i\hat  L_{e_a}$. Therefore, all supersymmetric eigenstates have exactly zero eigenvalue. One of the supersymmetric states is the steady-state (or zero-eigenvalue) total probability distribution (top differential form): the state of thermal equilibrium also known as ergodic zero~\footnote{In fact, each de Rahm cohomology class has one supersymmetric eigenstate -- otherwise the eigensystem~(\ref{eigensystem}) would not be complete on $\Omega$ and such situation would contradict the theory of pseudo-Hermitian operators~\cite{Mos023}. }. In Fig.~\ref{Fig_Spectrum} we schematically show the only three qualitatively different spectra that can emerge from the eigensystem~(\ref{eigensystem}). 

One of the key objects in the theory is the dynamical partition function (DPF), the trace of the finite-time SEO,
\begin{equation}
	Z_{tt'}= \text{Tr} \hat {\mathcal M}_{tt'} = \sum\nolimits_{n} e^{-\mathcal{E}_n(t'-t)}.\label{PartitionFunction}
\end{equation}
The DPF is a representative of periodic solutions of SDE. In fact, with little effort, it can be shown that in the long-time limit and for systems whose (stochastic) Lyapunov exponents are non-zero, the following inequality holds~\cite{Entropy}
\begin{eqnarray}
	\left. Z_{tt'} \right|_{t-t'\to\infty} \le \left\langle \# \{x| x=M_{t't}(x) \}\right\rangle = e^{S_{t}\times (t-t')},\label{Top_Entropy}
\end{eqnarray}
namely the DPF is bounded by an exponential growth of the number of fixed points of the SDE and whose rate is the constant, $S_{t}\ge0$, known in the the dynamical system theory as topological entropy.

We now have all the ingredients to discuss the relation between TSB and chaos. We first focus on its stochastic generalization and then on its deterministic properties. 

\subsection{Positive topological entropy}\label{SecIV}


The ground state of the system is the one that, in the long-time limit, contributes the most to the DPF~(\ref{PartitionFunction}). This is the eigenstate with the smallest real part of its eigenvalue. If this quantity is negative, then the ground state is non-supersymmetric because supersymmetric states have strictly zero eigenvalue. This situation is precisely the spontaneous breakdown of topological supersymmetry. This also implies that the topological entropy is positive in Eq.~(\ref{Top_Entropy}), which is typical of chaotic behavior.


\subsection{Sensitivity to initial conditions}\label{SecVI}


The stochastic generalization of sensitivity to initial conditions takes the form of infinite memory in response to external perturbations. A perturbation to the system at some moment $t_0$ can be viewed as the modification of the flow 
vector field in Eq.~(\ref{SDE}) by the addition of $\delta F=\delta(t-t_0)v$, with $v$ some (small) vector field. On the level of the evolution operator, Eq.~(\ref{H_SEO}), this corresponds to a perturbation of the type $\delta \hat H = \delta (t-t_0)\hat L_v$. In the long-time limit, the induced change in the partition function is then $\delta Z \propto \sum_{g} \langle g | \hat L_v | g \rangle$ with $g$ running over the ground states. This 
quantity is representative of the long-term memory if it is non-vanishing. This is indeed the case only if the ground states are non-supersymmetric because the Lie derivative, $\hat L_v$, is a $\hat d$-exact operator (according to Cartan's formula: $\hat{L}_v=[\hat d, \hat{\imath}_{v}]$). Since vanishing expectation values of all $\hat d$-exact operators is the key feature of supersymmetric states, this proves the statement that TSB implies sensitivity to initial conditions.

Note that the reverse is also true: if the DPF is responsive in the long-time limit to external perturbations at arbitrary times, or, in other words, if the system exhibits the ``butterfly effect'', then the ground state(s) is non-supersymmetric and the topological supersymmetry is spontaneously broken.


\subsection{Poincar\'e-Bendixson theorem}


The properties of the SEO eigensystem, Eq.~(\ref{eigensystem}), can be complemented with the observation that the topological supersymmetry can never broken by differential forms of zero-th and top degrees. For top degree forms, this follows from the understanding that top differential forms represent total probability distributions~\cite{Chaos}, and any such distribution tends to a stationary distribution in the long-time limit: the supersymmetric eigenstate with zero eigenvalue. The same property for zero-th order differential forms follows from the fact that they must have the same spectrum as top differential forms: $\textrm{spec}(\hat H^{(0)}) = \textrm{spec}(\hat H_{T}^{(D)})$, where $\hat H_T$ is the SEO of the time-reversed SDE. It then follows that the topological supersymmetry can only be spontaneously broken if $\dim X >2$. Only for these cases the properties of the spectrum of the SEO allow for a non-supersymmetric pair with negative real part of its eigenvalue. This is the stochastic generalization of the Poincar\'e-Bendixon theorem which states that deterministic chaos cannot occur for continuous dynamical systems in less than three dimensions~\cite{Vulpiani}.

\subsection{Non-integrability}\label{SecIII}  

Integrability of a deterministic flow in the sense of dynamical systems means that all \emph{global} unstable manifolds (GUMs) are well defined topological manifolds. For non-integrable flows, this is not the case. Instead, GUMs are self-folding fractal-like structures~\cite{E_Ruelle} that in the topological theory of chaos \cite{Gilmore} are approximately represented by ``branching manifolds''. 

The relation between integrability and topological supersymmetry is established via the concept of Poincar\'e duals of GUM~\cite{Topo-book,Entropy} --differential forms that are constant functions along the GUM and are $\delta$-function distributions, i.e., contain differentials, in the transverse directions. From the point of view of the SEO~(\ref{SEO}) in the deterministic limit, Poincar\'e duals of GUMs of integrable flow are the eigenstates with zero eigenvalues. These eigenstates are $\hat d$-closed, i.e., the operator $\hat d$ annihilates them. The operator $\hat d$ is (in cohomology) the operator version of the boundary 
operator (in homology)~\cite{Topo-book}, so that the $\hat d$-closeness of Poincar\'e duals of GUMs means that GUMs have no boundaries. Furthermore, if a GUM is non-trivial in homology, its Poincar\'e dual is nontrivial in the corresponding de Rahm cohomology and such eigenstate, which must be among the ground states of the model, is supersymmetric. In this manner, unbroken topological supersymmetry is equivalent, in the deterministic limit, to integrability.

From the complementary perspective of non-integrable flows, the Poincar\'e duals of GUMs are either not well defined, or they must be complemented by factors that will render them dependent on position on the GUM, which implies that they are not $\hat d$-closed (non-supersymemtric). In the assumption that GUMs, whether well-defined or not, are related to the ground states of the system, the topological supersymmetry must be broken spontaneously for non-integrable flows.

The Poincar\'e duals of GUMs represent the deterministic limit of the supersymmetric ground states when the topological supersymmetry is unbroken, whereas TSB must be associated with the stochastic generalization of non-integrability, namely stochastic chaos. Under conditions of stochastic integrability, the supersymmetric ground states can be roughly viewed as the stochastically-averaged Poincar\'e duals of GUMs. In fact, the reason why (non)integrability admits a stochastic generalization at all is because it involves linear objects -- Poincar\'e duals of GUMs are elements of the exterior algebra, which is a linear space -- and linear objects can be unambiguously averaged over the configurations of the noise. 

The same can be said about the other features discussed in this section. This is in contrast with the properties of set-theoretic chaos that we set to discuss in the next section. 


\section{Set-Theoretic Chaos}
\label{Sec:II}

Let us first note that, as we have discussed in the introduction, STC (in particular, topological mixing and dense periodic orbits) implies sensitivity to initial conditions (long-range temporal order). This long-range order can be interpreted as the result of the Goldstone theorem~\cite{Goldstone} applied to topological supersymmetry. This implies TSB in the deterministic limit, which is equivalent, as discussed in the previous Section, to non-integrability of the flow. 

The reverse must also be true. After all, non-integrability and STC for flows are two alternative mathematical views of the same phenomenon. Therefore, STC for flows and nonintegrability must be equivalent under general conditions, at least for those that cover physically realizable systems. 

On the contrary, unlike non-integrability, whose stochastic generalization is the TSB as discussed in the previous Section, STC and its trajectory-based properties -- topological transitivity/mixing and dense periodic orbits -- {\it do not admit} a stochastic generalization. 

To see this more clearly, consider the following simple stochastic dynamical system: stochastic evolution on (an open subset of) $\mathbb{R}^D$ with zero flow vector field and with ``Euclidean noise'', i.e., $e^i_a(x)=\delta^i_a$. Equation~(\ref{SDE}) is then, 
\begin{eqnarray}
	\dot x(t) = (2\Theta)^{1/2}\xi(t).\label{SDE-simple}
\end{eqnarray}  
In this case, the infinitesimal SEO~(\ref{H_SEO}) is just the Laplace operator in Euclidean space, $\hat H=-\sum_{a=1}^D(\partial/\partial x^a)$. The Green's function of the stochastic evolution equation~(\ref{SDE-simple}) is easily found:
\begin{eqnarray}
	&(\partial/\partial \tau + \hat H)G(\tau,x-x') = \delta^D (x-x'),\\
	&G(\tau,x-x') \propto \theta(\tau)\tau^{-D/2}e^{-(x-x')^2/(2\tau)},\label{GreenFunc}
\end{eqnarray}
with $\tau =\Theta t$ the reduced time, and $\theta$ the Heavyside step function. This Green's function~(\ref{GreenFunc}) represents the law of evolution of the coordinates on $X$, the bosonic sector of the wavefunctions of the model. 

In the path-integral representation of the theory, the Green's function is simply the weighted sum over all possible trajectories starting at $x$ and ending at $x'$ at a later moment of time $\tau$,
\begin{eqnarray}
	&G(\tau,x-x') \propto \iint_{x(0)=x,x(\tau)=x'} Dx(t) P(\xi),\label{path}
\end{eqnarray}
where the weight of each trajectory is the probability  (\ref{Noise_Probability}) of the noise configuration needed for the realization of this particular trajectory, $\xi(t) = \dot x(t)/(2\Theta)^{1/2}$, as follows from Eq.~(\ref{SDE-simple}).

Eqs.~(\ref{GreenFunc}) and (\ref{path}) show that no matter how far $x$ and $x'$ are from each other, and how short the time of the propagation $\tau$ is, there are always trajectories that connect these two points. This means, in particular, that Eq.(\ref{SDE-simple}) possesses the topological mixing property of the STC. 

This is indeed not surprising. Just like in quantum dynamics, in stochastic dynamics all trajectories are possible and the trajectory-based properties of STC are trivially satisfied by any stochastic model, whether chaotic (in the sense of TSB) or not. One way of interpreting this observation is as follows. STC highlights the ``noisy aspect'' of chaotic behavior and since noise is an intrinsic property of all stochastic models, chaotic or not, the latter ones automatically satisfy these conditions. 

One advantage of STC is its applicability to step-like (discrete-time) dynamics where chaotic behavior can exist even for phase spaces of dimensionality less than three~\cite{Vulpiani}. From the perspective of topological supersymmetry, this can be interpreted as the ability of the step-like dynamics to break topological supersymmetry {\it explicitly}, i.e., the evolution operator~(\ref{H_SEO}) does not commute with $\hat d$ if the map is not continuous. 

It must be noted, however, that spontaneous and explicit breakdowns of a symmetry are {\it qualitatively different phenomena}. For instance, the Goldstone's theorem~\cite{Goldstone} is applicable only to spontaneous symmetry breaking and \emph{not} to the explicit symmetry breaking. In the context of chaotic dynamics, this means that explicit supersymmetry breaking may not necessarily lead to the butterfly effect. In other words, the attempt to unify the concept of chaos for both  continuous-time {\it and} step-like dynamics within the {\it same} phenomenon of supersymmetry breaking, both spontaneous {\it and} explicit, may be a reasonable, but not (in our view) an optimal, solution.    



\section*{Conclusions}\label{Conclusions}


It was previously shown that the spontaneous breakdown of a topological supersymmetry can be associated with the stochastic generalization of non-integrability in the sense of dynamical systems and a few other features related to chaotic dynamics such as positive topological entropy, sensitivity to initial conditions, and the Poincar\'e-Bendixson theorem. In this work, we have strengthened and completed this picture by discussing its relation to the features associated with set-theoretic chaos 
--topological transitivity/mixing and dense periodic orbits-- that are widely considered fundamental to declare a system chaotic. 

We have shown that set-theoretic chaos is {\it not} generalizable to stochastic dynamics, and discussed that, in the deterministic case, it must be equivalent to  non-integrability, which is the deterministic limit of the {\it spontaneous} breakdown of topological supersymmetry. 

Therefore, since all dynamical systems in Nature are truly stochastic, topological supersymmetry breaking 
appears to be the most general and natural definition of chaos from a physical point of view, whether one considers its stochastic generalization or 
its deterministic limit.

Topological supersymmetry is a property of {\it any} continuous-time dynamical system described by stochastic or deterministic (partial) differential equations. Of course, only some dynamical systems have this supersymmetry broken spontaneously. Nevertheless, the universality of this supersymmetry unifies the phenomenology of chaos under the simple, general physical principle that many ``orders'' in Nature emerge from the phenomenon of spontaneous symmetry breaking. 

Therefore, this topological supersymmetry breaking picture leads us to a diametrically opposite view on the nature of dynamical chaos. In contrast to the meaning suggested by its name, chaos in continuous-time dynamical systems should be really viewed as the {\it ordered}, {\it low-symmetry phase} of dynamical systems that manifest it. In the present case, the (long-range) order is temporal. We therefore suggest the word {\it chronotaxis} as a better representation of this phenomenon. 

We finally note that for discrete-time (step-like) dynamics, topological supersymmetry can be broken {\it explicitly}. The physics associated with an 
explicit breakdown of a symmetry --as opposed to a spontaneous one-- is completely different. For one, Goldstone theorem does not apply, hence the 
butterfly effect typical of chaotic dynamics, does not necessarily emerge. It is true that discrete-time dynamics has a more mathematical, rather than physical, importance, but it would nonetheless be interesting to see if another, general physical phenomenon, in the spirit of what we have discussed in this work, holds also for this case.   

Irrespective, we note that the phase diagram at the 
border of chaos in the presence of noise is still a poorly understood topic, and we expect interesting new phenomena to emerge in that region. We then hope this work will motivate further studies in this direction.

{\it Acknowledgments --} M.D. acknowledges partial support from the Center for Memory and Recording Research at UCSD. I.V.O. would like to thank K. L. Wang for discussions.

\bibliographystyle{apsrev4-1}
\bibliography{SUSYref}
\end{document}